\documentclass{article}
\usepackage{graphicx} % Required for inserting images
\usepackage{authblk}

\title{LIME - a gas TPC prototype for directional Dark Matter search for the CYGNO experiment}

\author[1]{Fernando	Domingues Amaro} 
\author[2,3]{Elisabetta Baracchini} 
\author[4]{Luigi Benussi} 
\author[4]{Stefano Bianco}
\author[4]{Cesidio Capoccia}
\author[4,5]{Michele Caponero} 
\author[6]{Danilo Santos Cardoso} 
\author[7]{Gianluca Cavoto} 
\author[2,3]{Andr\'e Cortez} 
\author[9, 12]{Igor Abritta Costa} 
\author[4]{Emiliano Dan\'e}
\author[2,3]{Giorgio Dho}
\author[2,3]{\underline{Flaminia Di Giambattista}}
\author[8]{Emanuele Di Marco}
\author[8]{Giulia D'Imperio}
\author[8]{Francesco Iacoangeli}
\author[6]{Herman Pessoa Lima J\'unior}
\author[10]{Guilherme	Sebastiao Pinheiro	Lopes}
\author[4]{Giovanni Maccarrone}
\author[1]{Rui Daniel Passos Mano}
\author[11]{Robert Renz Marcelo Gregorio}
\author[2,3]{David Jos\'e Gaspar Marques}
\author[4]{Giovanni Mazzitelli}
\author[11]{Alasdair	Gregor	McLean}
\author[7,8]{Andrea Messina}
\author[1]{Cristina	Maria Bernardes	Monteiro}
\author[10]{Rafael	Antunes	Nobrega}
\author[10]{Igor Fonseca Pains}
\author[4]{Emiloano Paoletti}
\author[4]{Luciano Passamonti}
\author[8]{Sandro Pelosi}
\author[9,12]{Fabrizio Petrucci}
\author[7,8]{Stefano Piacentini}
\author[4]{Davide Piccolo}
\author[4]{Daniele Pierluigi}
\author[8]{Davide Pinci}
\author[2,3]{Atul Prajapati}
\author[8]{Francesco Renga}
\author[1]{Rita	Joanna da Cruz	Roque}
\author[4]{Filippo Rosatelli}
\author[4]{Alessandro Russo}
\author[1]{Joaquim	Marques Ferreira	dos Santos}
\author[4,13]{Giovanna Saviano}
\author[11]{Neil	John Curwen	Spooner}
\author[4]{Roberto Tesauro}
\author[4]{Sandro Tommasini}
\author[2,3]{Samuele Torelli}

\affil[1]{LIBPhys; Department of Physics; University of Coimbra; 3004-516 Coimbra; Portugal}%Department and Organization

\affil[2]{Gran Sasso Science Institute; 67100; L'Aquila; Italy}%Department and Organization

\affil[3]{Istituto Nazionale di Fisica Nucleare; Laboratori Nazionali del Gran Sasso; 67100; Assergi; Italy}%Department and Organization

\affil[4]{Istituto Nazionale di Fisica Nucleare; Laboratori Nazionali di Frascati; 00044; Frascati; Italy}%Department and Organization

\affil[5]{ENEA Centro Ricerche Frascati; 00044; Frascati; Italy}%Department and Organization

\affil[6]{Centro Brasileiro de Pesquisas F\'isicas; Rio de Janeiro 22290-180; RJ; Brazil}%Department and Organization

\affil[7]{Dipartimento di Fisica; Universit\`a La Sapienza di Roma; 00185; Roma; Italy}%Department and Organization

\affil[8]{Istituto Nazionale di Fisica Nucleare; Sezione di Roma; 00185; Rome; Italy}%Department and Organization

\affil[9]{Dipartimento di Matematica e Fisica; Universit\`a Roma TRE; 00146; Roma; Italy}%Department and Organization

\affil[10]{Universidade Federal de Juiz de Fora; Faculdade de Engenharia; 36036-900; Juiz de Fora; MG; Brasil}%Department and Organization

\affil[11]{Department of Physics and Astronomy; University of Sheffield; Sheffield; S3 7RH; UK}

\affil[12]{Istituto Nazionale di Fisica Nucleare; Sezione di Roma Tre; 00146; Rome; Italy}%Department and Organization

\affil[13]{Dipartimento di Ingegneria Chimica; Materiali e Ambiente; Sapienza Universit\`a di Roma; 00185; Roma; Italy}

\begin{document}

\maketitle

\begin{abstract}

The CYGNO experiment aims at the development of a large gaseous TPC with GEM-based amplification and an optical readout by means of PMTs and scientific CMOS cameras for 3D tracking down to O(keV) energies, for the directional detection of rare events such as low mass Dark Matter and solar neutrino interactions.
The largest prototype built so far towards the realisation of the CYGNO experiment demonstrator is the 50 L active volume LIME, with 4 PMTs and a single sCMOS imaging a 33$\times$33 cm\textsuperscript{2} area for 50 cm drift, that has been installed in underground Laboratori Nazionali del Gran Sasso in February 2022.
We will illustrate LIME performances as evaluated overground in Laboratori Nazionali di Frascati by means of radioactive X-ray sources, and in particular the detector stability, energy response and energy resolution. We will discuss the MC simulation developed to reproduce the detector response and show the comparison with actual data. We will furthermore examine the background simulation worked out for LIME underground data taking and illustrate the foreseen expected measurement and results in terms of natural and materials intrinsic radioactivity characterisation and measurement of the LNGS underground natural neutron flux. The results that will be obtained by underground LIME installation will be paramount in the optimisation of the CYGNO demonstrator, since this is foreseen to be composed by multiple modules with the same LIME dimensions and characteristics.
\end{abstract}

%% \linenumbers

%% main text
\section{The CYGNO experiment}
\label{sec:sample1}
The identification of the nature of Dark Matter (DM) is one of the most compelling tasks for fundamental physics today. A well motivated DM candidate are weakly interacting massive particles (WIMPs), which could interact with the ordinary matter of a target on Earth producing a nuclear recoil (NR). A promising approach is to look for the directional signature of low energy (less than 100 keV) NRs, whose angular distribution should point to the direction of the Cygnus constellation \cite{mayet}. 
The goal of the CYGNO experiment is the direct detection of DM, with the use of a gaseous time projection chamber (TPC) with an optical readout equipped with a triple-GEM (Gas Electron Multiplier) amplification stage, operated at atmospheric pressure \cite{cygno}. The ionization electrons produced by the interaction of charged particles within the gas volume are drifted by an electric field towards the triple GEM stack, where they are amplified. The secondary scintillation light produced in the electron avalanche is detected by scientific CMOS-based cameras (sCMOS), which capture the light emission projected on the GEM plane, and Photo Multiplier Tubes (PMTs) to determine the component along the drift direction. The He:CF4 gas mixture in 60/40 proportion provides a low energy threshold and a high scintillation yield. Several prototypes were built and are being used for R\&D studies towards the final goal of a O(1 m\textsuperscript{3}) CYGNO demonstrator with a modular design, which will be installed underground at Laboratori Nazionali del Gran Sasso (LNGS)  of INFN to prove the feasibility of the approach.

\section{The LIME prototype}
The 50 L LIME detector is the largest prototype we built so far, and it matches the dimension of one basic module of the future CYGNO demonstrator. Triple thin GEMs with an area of 33$\times$33 cm\textsuperscript{2} amplify the charge produced in the 50 cm long drift region. The high granularity Hamamatsu ORCA Fusion sCMOS and 4 PMTs located at the readout area's corners detect the secondary scintillation light. % produced by the amplified charge.  
The camera has 2304$\times$2304 pixels, each imaging an area of about 160$\times$160 \textmu m\textsuperscript{2}, with a low noise of about 1 photon per pixel, and a large quantum efficiency of 80\% at 600 nm, which nicely matches the spectral emission of our mixture. %The optical technique allows the readout sensors to be decoupled from the sensitive volume, preventing gas contamination.
The employment of the camera and PMTs together for 3D track reconstruction not only makes the detector sensitive to the direction of the events, but also allows for the fiducialization of the sensitive volume, lowering the impact of the background on the detector's sensitivity reach even further.
LIME was tested using multiple X-ray sources at the INFN Laboratori Nazionali di Frascati (LNF). For a month, the detector ran in stable conditions, with a measured rate of current spikes less that 2.7 per hour, confirming the possibility of underground continuous operation. The energy resolution was measured to be around 14\% across the 50 cm drift length using a \textsuperscript{55}Fe source generating 5.9 keV X-rays positioned on top of the sensitive volume at varying distances from the GEM plane. An energy threshold of 0.5 keV was set to ensure a maximum rate of 10 fake \textsuperscript{55}Fe events per year due to noise. A multivariate regression analysis, taking the position of the track and various shape parameters as inputs, is under development in order to improve the energy resolution. The preliminary results show an energy resolution better than 10\% at 5.9 keV. To evaluate the detector's response, multiple radioactive X-ray sources with energies ranging from 3.7 keV (from Ca) to 47 keV (from Tb) were employed, and a linear behavior was observed across the whole energy range.

\section{Monte Carlo simulation of the tracks}
To reproduce the electronic recoil (ER) and NR tracks seen by the sCMOS, a Monte Carlo (MC) simulation of the detector's response is being performed. First, a MC simulation of the ionization energy deposition as a function of 3D position is done using GEANT4 \cite{GEANT4} for the ERs and SRIM \cite{SRIM} for the NRs. Various processes (such as ionization e\textsuperscript{-} yield, diffusion along the drift region and inside the GEMs, charge amplification, photon yield and collection efficiency, absorption in the gas, gain saturation) are considered, and an image is produced, which is then analyzed using an intensity-based DBSCAN clustering algorithm \cite{dbscan}, which we also use for the analysis of the real images. The light integral and the track dimension of the simulated 5.9 keV ERs was compared to \textsuperscript{55}Fe data, showing a good agreement across different GEM gain values. In the linearity study, a preliminary comparison of X-ray data and the MC simulation reveals an agreement within 10\%. The study of the discrimination between NRs and ERs is ongoing in order to improve the background rejection capabilities of our approach, which was previously measured with a smaller 7 L prototype \cite{lemon}.

\section{Outlook}
In February 2022, LIME was installed underground at LNGS to be tested under the conditions of the future CYGNO experiment. These measurements will allow us to characterize the performance of the detector in low radioactivity and low pile-up configurations, measure the actual background, and validate the MC simulation, as well as test the gas system and all the data analysis tools developed for 3D track reconstruction and background rejection. The expected background was simulated, taking into account both the intrinsic radioactivity of the detector materials and the natural ambient gamma and neutron flux. The application of fiducial cuts to our sensitive volume showed that the radioactivity induced background events can be reduced by 96\%. The shielding design was optimized through dedicated MC simulations for different phases of the upcoming data collection underground, and it comprises the use of copper to shield the detector from gammas and water tanks to shield it from neutrons. Before the installation of the water shielding, a spectral measurement of the fast neutron flux underground will be done, which will provide valuable information for all rare events search experiments running underground at LNGS. 
The measurement will be done with LIME, which can detect the NRs produced by the elastic scattering of fast environmental neutrons, whose spectrum can be retrieved by deconvolving the measured NR spectrum with the response of the detector.
LIME is a critical step toward the development of the CYGNO demonstrator, which will demonstrate the scalability of this approach to large volumes.

\section*{Acknowledgments}
This project has received fundings under the European Union's Horizon 2020 research and innovation programme from the European Research Council (ERC) grant agreement No 818744. This project is supported by the Italian Ministry of Education, University and Research through the project PRIN: Progetti di Ricerca di Rilevante Interesse Nazionale "Zero Radioactivity in Future experiment" (Prot. 2017T54J9J). We want to thank General Services and Mechanical Workshops of Laboratori Nazionali di Frascati (LNF) and Laboratori Nazionali del Gran Sasso (LNGS) for their precious work and L. Leonzi (LNGS) for technical support.

% \bibliographystyle{plain} 
% \bibliography{biblio}

\end{document}